\documentclass[12pt,a4paper]{article}
\usepackage{latexsym}
\let\oldref=\ref  \def\ref#1{(\oldref{#1})}

\newcommand{\bee}{\begin{eqnarray}}
\newcommand{\eee}{\end{eqnarray}}
\newcommand{\be}{\begin{equation}}
\newcommand{\ee}{\end{equation}}

\newcommand{\f}{\frac}
\newcommand{\p}{\partial}
\newcommand{\half}{\frac{1}{2}}
\newcommand{\ga}{\alpha}

\newcommand{\gb}{\beta}
\newcommand{\gga}{\gamma}
\newcommand{\gd}{\delta}

\newcommand{\M}{{\cal M}}

\begin{document}
\begin{flushright}
\vspace{1mm}
 FIAN/TD/07--03\\
\vspace{-1mm}
\end{flushright}\vspace{1cm}

\begin{center}
{\large\bf  On $Sp\,(2M)$ Invariant Green Functions} \vglue 0.6 true
cm \vskip0.8cm {M.A.~Vasiliev and V.N.~Zaikin}
\vglue 0.3  true cm

I.E.Tamm Department of Theoretical Physics, Lebedev Physical Institute,\\
Leninsky prospect 53, 119991, Moscow, Russia 
\vskip1.3cm \end{center}

\begin{abstract}
Explicit form of two-point  and three-point $Sp\,(2M)$  invariant
Green functions is found.
\end{abstract}

\section{Introduction}
As was originally pointed out by Fronsdal \cite{Fr}, infinite
towers of massless unitary representations of all spins
 of the $AdS_4$ symmetry $sp(4)\sim o(3,2)$
exhibit higher symplectic symmetry $sp(8)$ which extends usual
conformal symmetry $su(2,2)$. It was also pointed out in \cite{Fr}
that the minimal space-time in which $sp(8)$ acts geometrically is
ten dimensional. It can be realized as the space of Lagrangian
planes \cite{Fr} as well as the coset space $Sp(8)/P$ where $P$ is
an appropriate parabolic subgroup of $Sp(8)$ \cite{Mar}. The local
coordinates of this generalized space-time $\M_4$  are symmetric
matrices $X^{\ga\gb}=X^{\gb\ga}$, where $\ga,\gb\ldots =1,2,3,4$
are four dimensional Majorana spinor indices. In this paper we
will not distinguish between $\M_4$ and its large cell $R^{\half
M(M+1)}$. All integer spin massless bosons and half-integer
massless fermions in four dimensions are described by a single
scalar field $\Phi(X)$ and spinor field $\Phi_\ga (X)$ in $\M_4$,
respectively.

Infinitesimal $Sp(8)$ transformations are \cite{Mar}
\begin{eqnarray}
\label{trscal} \delta\Phi^{d}(X)=\left(\varepsilon^{\alpha\beta}
\frac{\partial}{\partial
X^{\alpha\beta}}+{}d{}\varepsilon^{\alpha}{}_{\alpha}+
2\varepsilon^{\alpha}{}_{\beta}X^{\beta\gamma}\frac{\partial}{\partial
X^{\alpha\gamma}}\nonumber \right.\\ \left. -
\varepsilon_{\alpha\beta}\left[{}d{}X^{\alpha\beta}+
X^{\alpha\gamma}X^{\beta\eta}\frac{\partial}{\partial
X^{\gamma\eta}} \right] \right) \Phi^{d}(X),\label{gen Phi}
\end{eqnarray}
\begin{eqnarray} \label{trspin}
\delta\Psi^{\Delta}_{\rho}(X)=\left(\varepsilon^{\alpha\beta}
\frac{\partial}{\partial
X^{\alpha\beta}}+{}\Delta{}\varepsilon^{\alpha}{}_{\alpha}+
2\varepsilon^{\alpha}{}_{\beta}X^{\beta\gamma}\frac{\partial}{\partial
X^{\alpha\gamma}}\nonumber \right.\\ \left. -
\varepsilon_{\alpha\beta}\left[{}\Delta{}X^{\alpha\beta}+
X^{\alpha\gamma}X^{\beta\eta}\frac{\partial}{\partial
X^{\gamma\eta}} \right] \right) \Psi^{\Delta}_{\rho}(X)+
(\varepsilon^{\beta}{}_{\rho}-\varepsilon_{\eta\rho}X^{\eta\beta})
\Psi^{\Delta}_{\beta}(X) ,\label{gen Psi} \end{eqnarray} where
$\varepsilon^{\alpha\beta}$, $\varepsilon^{\alpha}{}_\beta$, and
$\varepsilon_{\alpha\beta}$ are $X$ - independent parameters of,
respectively, generalized translations, Lorentz transformations
along with dilatations, and special conformal transformations. $d$
and $\Delta$ are conformal weights of the fields $\Phi^{d}$ and
$\Psi^{\Delta}_{\rho}$, respectively.

The $sp(8)$ invariant form of the free massless field equations in
$\M_4$ is \cite{BHS} \be \label{oscal} \Big ( \f{\p^2}{\p
X^{\ga\gb} \p X^{\gga\gd}} - \f{\p^2}{\p X^{\ga\gga} \p
X^{\gb\gd}}\Big ) \Phi(X) =0 \ee and \be \label{ofer} \f{\p}{\p
X^{\ga\gb}} \Phi_\gga(X)       - \f{\p}{\p X^{\ga\gga}}
\Phi_\gb(X)      =0\,. \ee These equations are invariant under the
$sp(8)$ transformations \ref{trscal} and \ref{trspin} with the
canonical conformal dimensions $d= \Delta = \half$. That these
equations indeed describe four dimensional higher spin dynamics
and are invariant under $sp(8)$ symmetry was shown in \cite{BHS}
based on the so called unfolded form of the higher spin equations
in the form of certain covariant constancy conditions. Note that
in \cite{BLS} a $sp(8)$ world-line particle model was considered
giving rise to the first-class constraints which upon Dirac
quantization give rise to the equations analogous to those of
\cite{BHS}. The equations \ref{oscal} and \ref{ofer} make sense
for any even number $M$ of values taken by the indices $\ga,\gb$
and remain  $sp(2M)$ invariant. As argued in \cite{BLS,Mar}, the
related dynamical systems are expected to be related with
conformal higher spin systems in six and ten dimensions for $M=8$
and 16, respectively. We therefore consider in this paper any even
$M$. Note that the form of the transformation laws \ref{trscal}
and \ref{trspin} is $M$-independent.

It is yet an open question whether some nontrivial higher spin theories
exist which exhibit higher symplectic symmetries in the unbroken
phase beyond the free field level. There is some evidence, however,
that spontaneously broken symmetries of this type  play a role in
supergravity \cite{West}. Also spontaneously broken higher symplectic
symmetries appear in the nonlinear higher spin field equations as
formulated in \cite{more}. If there is a
phase in which higher symplectic symmetries are unbroken
it should describe some higher spin theory because
irreducible representations of higher symplectic symmetries form
infinite towers of massless fields.

The aim of this  note is to analyze general restrictions on $sp(2M)$
invariant Green functions. Following  standard methods of
conformal field theory along the lines of \cite{Pol}  (see also
\cite{Fra-Pal} and references therein) we derive the forms of
two-point and three-point functions
invariant under the $Sp~(2M)$
transformations \ref{trscal} and \ref{trspin}. The final results
are
\begin{equation}\label{res-Phi-2}
\langle\Phi^{d_{1}}(X_{1})\Phi^{d_{2}}(X_{2})\rangle{}
={}C_{\Phi\Phi}(\det|X_{1}-X_{2}|)~^{-d_{1}}\,,\qquad
 d_{1} = d_{2} \,
\end{equation}
\begin{equation}\label{res-Psi-2}
\langle\Psi^{\Delta_{1}}_{\rho_{1}}(X_{1})\Psi^{\Delta_{2}}_{\rho_{2}}(X_{2})\rangle
=C_{\Psi\Psi}\,(X_{1}-X_{2})_{\rho_{1}\rho_{2}}\left(\det|X_{1}-X_{2}|\right)^{-\Delta
_{1}}\,,\qquad \Delta_{1} = \Delta_{2} \,
\end{equation}

\begin{equation}
\begin{array}{c}
\label{res-Phi-3}
\langle\Phi^{d_{1}}(X_{1})\Phi^{d_{2}}(X_{2})\Phi^{d_{3}}(X_{3})\rangle =
C_{\Phi\Phi\Phi}(\det|X_{1}-X_{3}|)^{-\half(d_{1} + d_{3} -
d_{2})}
\\
\\
\times(\det|X_{2}-X_{3}|)^{-\half(d_{2} + d_{3} - d_{1}) }
(\det|X_{1}-X_{2}|)^{-\half(d_{1} + d_{2} - d_{3}) },
\end{array}
\end{equation}

\begin{equation}
\begin{array}{c}
\label{res-psi-3}
\langle\Psi^{\Delta_{1}}_{\rho_{1}}(X_{1})\Psi^{\Delta_{2}}_{\rho_{2}}(X_
{2})\Phi^{d}(X_{3})\rangle=C_{\Psi\Psi\Phi}\,(X_{1}-X_{2})_{\rho_{1}\rho_{2}}
{}\left(\det|X_{1}-X_{3}|\right)^{-\frac{\Delta_{1}-\Delta_{2}+d}{2}}
\\
\\
\times\left(\det|X_{2}-X_{3}|\right)^{-\frac{\Delta_{2}-\Delta_{1}+d}{
2}}
\left(\det|X_{1}-X_{2}|\right)^{-\frac{\Delta_{1}+\Delta_{2}-d}{2}}\,.
\end{array}
\end{equation}
(Green functions containing odd numbers of fermions $\Psi$ vanish.)

This form of $sp(2M)$ invariant
Green functions is analogous to that of usual conformal
Green functions with the invariant intervals replaced by the
determinants of matrix coordinates.  It is tempting to speculate that
the obtained results should admit an extension to higher Green
functions and/or multispinor fields.  The fact that $sp(2M)$ invariance
allows nonzero three--point Green functions we interpret as an
indication that nontrivial generalized
conformal theories with higher symplectic symmetries may exist
beyond the free field level.

In the rest of this letter we sketch the proof of the formulae
\ref{res-Phi-2}-\ref{res-psi-3}.

\section{Two--point functions}

As a consequence of~\ref{gen Phi} the $sp(2M)$ two point function
$$G(X_{1},X_{2}) =
\left\langle\Phi^{d_{1}}(X_{1})\Phi^{d_{2}}(X_{2})\right\rangle$$
has to satisfy the following conditions
\begin{eqnarray}
\varepsilon^{\alpha\beta}\left\{\frac{\partial}{\partial
X^{\alpha\beta}_{1}} + \frac{\partial}{\partial
X^{\alpha\beta}_{2}}\right\}G(X_{1},X_{2})&=&0 \label{P-2-Phi}\,,\\
\left\{\varepsilon^{\alpha}{}_{\alpha}\left(d_{1}+d_{2}\right)+
2\varepsilon^{\alpha}{}_{\beta}\left(X_{1}^{\beta\gamma}\frac{\partial}
{\partial X_{1}^{\alpha\gamma}} +
X_{2}^{\beta\gamma}\frac{\partial}{\partial
X_{2}^{\alpha\gamma}}\right)\right\}G(X_{1},X_{2})&=&0
\label{D-2-Phi}\,, \\
\varepsilon_{\alpha\beta}\left\{{}d_{1}{}X_{1}^{\alpha\beta}+
X_{1}^{\alpha\gamma}X_{1}^{\beta\eta}\frac{\partial}{\partial
X_{1}^{\gamma\eta}} +{}d_{2}{}X_{2}^{\alpha\beta}+
X_{2}^{\alpha\gamma}X_{2}^{\beta\eta}\frac{\partial}{\partial
X_{2}^{\gamma\eta}}  \right\}G(X_{1},X_{2})&=&0 \label{K-2-Phi}\,.
\end{eqnarray} {}From \ref{P-2-Phi} it follows that \be \label{sol}
G(X_{1},X_{2})=F(Y)\,,\qquad Y=X_1 - X_2 \ee Eq. \ref{D-2-Phi}
then transforms to \begin{equation}\label{D'-2-Phi}
  \varepsilon^{\alpha}{}_{\alpha}\left(d_{1}+ d_{2}\right) +
  2{}\varepsilon^{\alpha}_{\beta}Y^{\beta\gamma}\frac{\partial}{\partial
  Y^{\alpha\gamma}}P(Y) = 0,
\end{equation}
where $P(Y){}={}\ln F(Y)$. Taking  into account
\begin{equation}\label{usful1}
  \frac{\partial}{\partial
  X^{\alpha\beta}}\det|X|{}={}X_{\alpha\beta}\det |X|\,,\qquad
 X_{\beta\gamma}X^{\gamma\alpha}{}={}\delta_{\beta}{}^{\alpha}\,,
\end{equation}
the general solution of this equation is
\begin{equation}\label{sol Phi-2}
 P(Y){}={}-\frac{d_{1}+d_{2}}{2}\ln (\det|Y|)+ C,
\end{equation}
i.e.,
\begin{equation}\label{sol'-Phi-2}
  G(X_{1},X_{2}){}={}C(\det|X_{1}-X_{2}|)~^{-\frac{d_{1}+d_{2}}{2}},
\end{equation}
where $C$ is an arbitrary constant. Substituting \ref{sol'-Phi-2}
into \ref{K-2-Phi} one gets
\begin{equation}\label{sol"-Phi-2}
(d_{1}-d_{2})\epsilon^{\alpha\beta}(X_{1}-X_{2})_{\alpha\beta}{}={}0,
\end{equation}
 i.e.  $G(X_{1},X_{2}) =
\left\langle\Phi^{d_{1}}(X_{1})\Phi^{d_{2}}(X_{2})\right\rangle$
can be nonzero only if $d_{1} = d_{2}$. As a result, we get the
scalar $sp(2M)$ invariant two-point function \ref{res-Phi-2}. For
the free field case of $d_1=d_2=\half$ this agrees with the free
Green function found in \cite{Mar}.

Consider now the fermionic two point function
$$
G_{\rho_{1}\rho_{2}}(X_{1},X_{2}) =
\langle\Psi^{\Delta_{1}}_{\rho_{1}}(X_{1})\Psi^{\Delta_{2}}_{\rho_{2}}(X_{
2})\rangle\,.
$$
Taking into account \ref{gen Psi}{}, the analysis analogous to the
scalar case (cf. Eqs.\ref{sol}-\ref{sol"-Phi-2}) gives $\Delta_{1}
= \Delta_{2}$ and
\begin{equation}\label{sol-psi-2}
G_{\rho_{1}\rho_{2}}(X_{1},X_{2}) =
\left(\det|Y|\right)^{-\Delta_{1}}P_{\rho_{1}\rho_{2}}(Y),
 \end{equation}
 where
 $P_{\rho_{1}\rho_{2}}(Y)$ is some function satisfying the equations
\begin{eqnarray} 2Y^{\beta\eta}\frac{\partial}{\partial
Y^{\alpha\eta}}P_{\rho_{1}\rho_{2}}(Y) +
\delta^{\beta}{}_{\rho_{1}}P_{\alpha\rho_{2}}(Y) +
\delta^{\beta}{}_{\rho_{2}}P_{\rho_{1}\alpha}(Y)= 0\label{D-psi-2}\,,\\
2Y^{\alpha\delta}Y^{\beta\eta}\frac{\partial}{\partial
Y^{\delta\eta}}P_{\rho_{1}\rho_{2}}(Y) +
\delta^{\alpha}{}_{\rho_{1}}Y^{\beta\delta}P_{\delta\rho_{2}}(Y) +
\delta^{\beta}{}_{\rho_{1}}Y^{\alpha\delta}P_{\delta\rho_{2}}(Y)=
0\label{K-psi-2}\,. \end{eqnarray} {}From \ref{D-psi-2},
\ref{K-psi-2} it follows that. $$
  \delta^{\alpha}{}_{\rho_{1}}Y^{\beta\delta}P_{\delta\rho_{2}}(Y)
  -
   \delta^{\beta}{}_{\rho_{2}}Y^{\alpha\delta}P_{\rho_{1}\delta}(Y
   )
  = 0
$$
and, therefore,
\begin{equation}\label{sol-Psi-2}
  P_{\rho_{1}\rho_{2}}(Y) = (Y)_{\rho_{1}\rho_{2}}P(Y)\,.
\end{equation}
Plugging  \ref{sol-Psi-2} into \ref{D-psi-2} and \ref{K-psi-2} we
find that $ P(Y) = const. $ Finally we obtain the expression ~\ref{res-Psi-2}
for the fermionic Green function. For the canonical dimensions
$\Delta_1=\Delta_2= \half$ one recovers the free fermionic Green
function of \cite{Mar}.

\section{Three--point functions}
As a consequence of~\ref{gen Phi} the Green function
$$G(X_{1},X_{2},X_{3}) =
\left\langle\Phi^{d_{1}}(X_{1})\Phi^{d_{2}}(X_{2})\Phi^{d_{3}}(X_{3})
\right\rangle$$ has to satisfy the following system of equations
\begin{eqnarray} \varepsilon^{\alpha\beta}\sum_{i=1}^{3}
\frac{\partial}{\partial
X_{i}^{\alpha\beta}}G(X_{1},X_{2},X_{3})&=&0\,,\label{P-3-Phi}\\
\left\{\varepsilon^{\alpha}_{\alpha}\sum_{i=1}^{3} d_{i} +
2\varepsilon^{\alpha}_{\beta}\sum_{i=1}^{3}
X_{i}^{\beta\gamma}\frac{\partial}{\partial
X_{i}^{\alpha\gamma}}\right\}G(X_{1},X_{2},X_{3})\,&=&0\,,\label{D-3-Phi}\\
\varepsilon_{\alpha\beta}\sum_{i=1}^{3}\left\{{}d_{i}{}X_{i}^{\alpha
\beta} +
X_{i}^{\alpha\gamma}X_{i}^{\beta\eta}\frac{\partial}{\partial
X_{i}^{\gamma\eta}}\right\}G(X_{1},X_{2},X_{3})&=&0\,.\label{K-3-Phi}
\end{eqnarray} {}From \ref{P-3-Phi} we obtain that
\begin{equation}\label{ind}
  G(X_{1},X_{2},X_{3}){}={}F(Y_{1},Y_{2}),
\end{equation}
 where we introduced new variables: \begin{equation}\label{new 2}
  Y_{1}^{\alpha\beta} = X_{1}^{\alpha\beta}-X_{3}^{\alpha\beta},
  \qquad Y_{2}^{\alpha\beta} =
  X_{2}^{\alpha\beta}-X_{3}^{\alpha\beta}\,.
\end{equation}
 Eq.\ref{D-3-Phi} then transforms to
\begin{equation}\label{D'-3-Phi}
  \varepsilon^{\alpha}_{\alpha}\sum_{i=1}^{3} d_{i} +
  2{}\varepsilon^{\alpha}_{\beta}\sum_{i=1}^{2}
  Y_{i}^{\beta\gamma}\frac{\partial}{\partial
  Y_{i}^{\alpha\gamma}}P(Y_{1},Y_{2}) = 0,
\end{equation}
where $P(Y_{1},Y_{2}) = \ln F(Y_{1},Y_{2})$. Its general solution
is
\begin{equation}\label{sol Phi-3}
  P(Y_{1},Y_{2}) = \ln\left[ (\det|Y_{1}|)^{-\frac{k_{1}}{2}}
  (\det|Y_{2}|)^{-\frac{k_{2}}{2}}(\det|Y_{1}-Y_{2}|)^{-\frac{k_3}{2}}
  \right] + R(Y_{1},Y_{2}),
\end{equation}
where $k_1$, $k_2$ and $k_3$ are such that
\begin{equation}\label{id 1}
  k_{1} + k_{2} + k_{3} = d_{1} + d_{2} + d_{3}
\end{equation}
and $R(Y_{1},Y_{2})$ satisfies the equation
\begin{equation}\label{D''-3-Phi}
  \left\{Y_{1}^{\beta\gamma}\frac{\partial}{\partial Y_{1}^{\alpha\gamma}}
  + Y_{2}^{\beta\gamma}\frac{\partial}{\partial
  Y_{2}^{\alpha\gamma}}\right\}R(Y_{1},Y_{2}) = 0
\end{equation}

Taking into account \ref{ind} and  \ref{D'-3-Phi} eq.\ref{K-3-Phi}
transforms to
\begin{equation}\label{K''-3-Phi}
  d_{1}Y_{1}^{\alpha\beta} + d_{2}Y_{2}^{\alpha\beta} +
  \left[Y_{1}^{\alpha\gamma}Y_{1}^{\beta\eta}\frac{\partial}{\partial
Y_{1}^{\gamma\eta}} +
Y_{2}^{\alpha\gamma}Y_{2}^{\beta\eta}\frac{\partial}{\partial
Y_{2}^{\gamma\eta}}\right]P(Y_{1},Y_{2}) = 0 \end{equation}
Plugging here the expression \ref{sol Phi-3} and choosing
 \begin{equation}\label{id 2}
  2d_{1} - k_{1} - k_{3} = 0,\quad  2d_{1} - k_{1} - k_{3} = 0,
\end{equation}
we get
\begin{equation}\label{K'''-3-Phi}
\left[Y_{1}^{\alpha\gamma}Y_{1}^{\beta\eta}\frac{\partial}{\partial
Y_{1}^{\gamma\eta}} +
Y_{2}^{\alpha\gamma}Y_{2}^{\beta\eta}\frac{\partial}{\partial
Y_{2}^{\gamma\eta}}\right]R(Y_{1},Y_{2}) = 0\,.
\end{equation}
It is easy to see that from \ref{D''-3-Phi} and \ref{K'''-3-Phi}
it follows
 $R(Y_{1},Y_{2}){}={}const.$
Thus \ref{res-Phi-3} is the general expression for a three-point
function.

For the three-point function
$$
G_{\rho_{1}\rho_{2}}(X_{1},X_{2},X_{3}) =
\langle\Psi^{\Delta_{1}}_{\rho_{1}}(X_{1})
\Psi^{\Delta_{2}}_{\rho_{2}}(X_{2})\Phi^{d}(X_{3})\rangle\,
$$
one obtains analogously from \ref{trscal} and \ref{gen Psi} the
expression \ref{res-psi-3}.
Also, one can see that the Green functions with odd numbers of
spinor fields vanish.



\vspace{1.5cm}

{\bf Acknowledgments}
\\
\\

This work is supported by grants  RFBR  No 02-02-17067, LSS No
1578.2003-2 and INTAS No 03-51-6346.


\end{document}